\documentclass[prl,twocolumn,letterpaper,showpacs,superscriptaddress]{revtex4}
\usepackage{graphicx,bm,amssymb,amsmath}

\def\be{\begin{equation}}
\def\fin{\end{equation}}

\def\T{{\sf T\kern-.45em T}}
\def\C{\kern.1em{\raise.47ex\hbox{$\scriptscriptstyle |$}}
             \kern-.40em{\sf C}}

\begin{document}
\title{Generic phase diagram of active polar films}
\date{\today}

\author{ R.Voituriez}
\affiliation{Physicochimie Curie (CNRS-UMR168), Institut Curie, Section
de Recherche,
26 rue d'Ulm 75248 Paris Cedex 05 France}
\author{J.F. Joanny}
\affiliation{Physicochimie Curie (CNRS-UMR168), Institut Curie, Section
de Recherche,
26 rue d'Ulm 75248 Paris Cedex 05 France}
\author{ J. Prost}
\affiliation{Physicochimie Curie (CNRS-UMR168), Institut Curie, Section
de Recherche,
26 rue d'Ulm 75248 Paris Cedex 05 France}
\affiliation{ E.S.P.C.I, 10 rue Vauquelin, 75231 Paris Cedex 05, France
}
\pacs{87.10.+e, 83.80.Lz, 61.25.Hq}

\begin{abstract}
We study theoretically the phase diagram of compressible active polar
gels such as the actin network of eukaryotic cells. Using
generalized hydrodynamics equations,  we perform a linear stability analysis
of the uniform states in the case of an infinite bidimensional active gel to obtain
the dynamic phase diagram of active polar films. We predict in
particular modulated flowing phases, and a macroscopic phase separation at high activity.
This qualitatively accounts for experimental observations of various active systems,
such as acto-myosin gels, microtubules and kinesins in vitro solutions, or swimming
bacterial colonies.

\end{abstract}
\maketitle

Active materials are a challenging class of systems driven out of
equilibrium by an internal or an external energy source. Examples of
active systems  are self--propelled particle assemblies in bacterial
colonies~\cite{gold,simh02}, or the membrane and the cytoskeleton of
eukaryotic cells~\cite{albe02}. The cell cytoskeleton is a network
of long filaments made by protein assembly, interacting with other
proteins~\cite{howa01} which can, among other things, crosslink or
cap the filaments. Motor proteins, myosins, kinesins or dyneins use
the chemical energy of Adenosinetriphosphate (ATP) hydrolysis to
"walk" along the filaments, and exert stresses that deform the
network~\cite{nede97}, leading to an active behavior. The active
character of the cytoskeleton plays a major role in most cell
functions such as intracellular transport, motility and cell
division.

The cell cytoskeleton has a rich and complex dynamical behavior
 \cite{nede97,krus00,lee01,kim03,live03}.
Self-organized  patterns, such as asters, vortices, and rotating
spirals, microscopic and macroscopic phase separations
("superprecipitation" \cite{japa} ) have been observed as a function
of motor and ATP concentrations in a thin film~\cite{nede97}. This
two-dimensional geometry gives for example a good description of the
thin lamellipodium of a cell spreading or moving on a substrate.
Some of these structures have been recently explained theoretically
\cite{krus04,menon}, but a full phase diagram of active polar films
is still missing.

The  passive visco-elastic properties of the cytoskeleton are
similar to that of  a physical gel made of the cross-linked
semi-flexible filaments. Recently, Kruse et al. \cite{krus04} have
proposed a generalized hydrodynamic theory to describe
macroscopically the {\it active} character of incompressible polar
gels, based on conservation laws and symmetry considerations. In
this letter, we use the generic model of Ref.\cite{krus04} to study
the stability of compressible active polar films. We perform a
linear stability analysis of the uniform states in the case of an
infinite two-dimensional geometry, and obtain the dynamic phase
diagram. Our results qualitatively account for the experimental
observations on various active systems, such as acto-myosin gels,
microtubules and kinesin solutions  in vitro or swimming bacterial
colonies. We choose the example of infinite acto-myosin films that
we consider as two-dimensional, and non interacting with the
environment. Such an example could be realized by a freely suspended
film such as those studied for liquid crystals. We consider the long
wavelength limit, and all variables are implicitly averaged over the
film thickness. The actin network has a local macroscopic
polarization given by a unitary vector ${\bf p}=(\cos
\theta,\sin\theta)$. This describes degenerate parallel boundary
conditions at the film interface, but also equivalently normal
boundary conditions with a splayed state. Since the film thickness
 can vary, the two--dimensional actin network is compressible, even though
  the three--dimensional material is not. The gel has
a weakly fluctuating density $c({\bf r}) =c_0 + \rho(\bf r)$. The
average density $c_0$ can be set to $1$ by rescaling the various
coefficients of the free energy given in Eq.\ref{free} below. As a
first approximation we neglect the interactions with the solvent and
use a one fluid model.

The free energy of the gel, up to quadratic order,
  couples the polarization ${\bf p}$ to the density fluctuation $\rho$ :

\begin{eqnarray}
F&=&\int dxdy\left[ \frac{K_1}{2}(\nabla\cdot {\bf  p} )^2 +\frac{K_3}{2}(\nabla\times{\bf  p})^2 +k\nabla{\bf  p} - \frac{1}{2}h_\parallel {\bf p}^2\right.\nonumber\\
  &  &  \left. +w\rho\nabla\cdot{\bf  p}+\frac{\beta}{2}(\nabla \rho )^2+\frac{\alpha}{2}\rho^2 \right]
\label{free}
\end{eqnarray}

The splay and bend elastic moduli are assumed to be equal
($K_1=K_3=K$)  for the  sake of simplicity. The Lagrange multiplier
$h_{||}$ enforces the constraint ${\bf p}^2=1$. The spontaneous
splay term leads to boundary terms that are irrelevant in the
infinite system size limit \cite{voit}. The variation of the free
energy with density is characterized by the positive compressibility
 $\alpha$ and by the positive coefficient $\beta$ associated to the density
fluctuation  correlation length; $w$ is a coupling constant between
density fluctuations and splay. The molecular field, conjugate to
the polarization is $h_\alpha=-\delta F/\delta p_\alpha$ with
coordinates $(h_\parallel,h_\perp)$ parallel and perpendicular to
the polarization. The free energy is similar to that of a
ferro-electric nematic liquid crystal when the order parameter ${\bf
n}$ does not have a fixed length \cite{pelc,blank}).

The gel motion is described by the two-dimensional velocity field
${\bf v}$. The strain rate tensor is
$u_{\alpha\beta}=(\partial_{\alpha}v_\beta+\partial_{\beta}v_\alpha)/2$,
and the vorticity tensor
$\omega_{\alpha\beta}=(\partial_{\alpha}v_\beta-\partial_{\beta}v_\alpha)/2$.
The  conservation equation of the gel is written as:
$\partial_t\rho+\partial_\alpha(1+\rho)v_\alpha=0$.

The gel is driven out of equilibrium by  continuous and homogeneous
input of energy, characterized by the chemical potential difference
$\Delta \mu$ between ATP and its hydrolysis products, which we
assume to be constant.

The gel dynamics is described by the linear
hydrodynamic equations for active polar gels of Ref.\cite{krus04}. The constitutive equations for the mechanical deviatory stress tensor $ \sigma_{\alpha\beta} $ and the rate of change of the polarization $\frac{D}{Dt}p_\alpha=\frac{\partial
p_\alpha}{\partial t}+(v_\gamma\partial_\gamma)p_\alpha
+\omega_{\alpha\beta} p_\beta$ read, at long time scales when
the gel behaves as a viscous liquid:
\begin{eqnarray}
2\eta u_{\alpha\beta}&=&\sigma_{\alpha\beta} + \zeta\Delta\mu p_\alpha p_\beta +
\bar\zeta\Delta\mu\delta_{\alpha\beta}-\frac{\nu}{2}(p_\alpha h_\beta+p_\beta h_\alpha)\nonumber\\
  &  &   - \bar\nu p_\gamma h_\gamma\delta_{\alpha\beta} + \frac{1}{2}(p_\alpha h_\beta-p_\beta h_\alpha)
\label{uab}\\
\frac{D p_\alpha}{D t}  &=&  \frac{1}{\gamma} h_\alpha + \lambda
p_\alpha \Delta\mu  - \nu u_{\alpha\beta}p_\beta-\bar\nu
u_{\beta\beta}p_\alpha \label{eq:dpdt}
\end{eqnarray}
 We  neglect here the geometric non-linearities introduced in \cite{krus04}. The rotational viscosity $\gamma$
and the coupling constants between
 flow and polarization $ \nu, \ {\bar \nu}$  are standard liquid crystal
parameters \cite{dege93}.
The active contributions to the
mechanical stress and to the rate of change of the polarization
are proportional to $\Delta \mu$ and are characterized by the
coefficients $\zeta$, ${\bar \zeta}$ and $\lambda$.
This set of constitutive equations is completed at low Reynolds
numbers by
the force balance: $\partial_\alpha
(\sigma_{\alpha\beta}-\Pi\delta_{\alpha\beta})=0$.
Locally, there are two forces acting on the gel, the deviatory
stress tensor $\sigma_{\alpha\beta}$ and the pressure $\Pi=\frac{\delta F}{\delta\rho}=w\nabla\cdot{\bf
p}+\alpha\rho-\beta\Delta\rho$.

It is observed experimentally \cite{kaes}, and predicted by
1-dimensional models \cite{krus00} that the overall effect of myosin
II motors on actin solutions is contractile. This corresponds to
negative values of both $\zeta$ and ${\bar \zeta}$. The molecular
motors also have an effect on the rate of change of the polarization
described by $\lambda$; if  $\lambda>0$  the polarization is
enhanced. This seems to be observed experimentally (zipping effect
in Ref.\cite{sack}). We consider here only ordered polar phases with
a unitary polarization vector ${\bf p}$, and we ignore this active
coupling  (setting $\lambda=0$) for simplicity  \cite{lambda}.

In order to discuss the accessible steady states of an infinite
active gel film, we first study the states of uniform polarization
and velocity gradient. These asymptotic states are non--equilibrium
states, and cannot therefore  be obtained by minimizing a free
energy functional. In a two-dimensional geometry, the hydrodynamic
equations of motion give 11 scalar equations for the 11 independent
variables ${\bf x}\equiv\theta, \rho, h_{\alpha}, v_{\alpha},
\sigma_{\alpha\beta}, \pi$. These equations of motion have two types
of homogeneous steady states: a static state where the velocity
gradient $u_{\alpha\beta}$ vanishes and the polarization is uniform
and oriented in a direction $\theta_0$; and a flowing state with a
finite velocity gradient and a uniform polarization.  In two
dimensions, the most general steady flow with a constant velocity
gradient is a superposition of two simple shear flows in two
perpendicular directions $x$ and $y$. The only non vanishing
components of the velocity gradient are $\partial_x v_y$ and
$\partial_y v_x$. We consider here for simplicity only one component
shear flows for which $\partial_y v_x=0$. The velocity is then along
the $y$ direction and the polarization angle $\theta_\nu$ is such
that $\cos(2\theta_\nu)=1/\nu$ (we assume $|\nu|>1$). This flowing
state is the analog, for an infinite compressible gel, of the
flowing state obtained in a confined geometry in \cite{voit} and to
the rotating spirals in a cylindrical geometry described in
\cite{krus04}. It confirms the possibility of obtaining spontaneous
flows in polar active materials.

The stability of the two homogeneous states is studied by introducing a small
perturbation at point ${\bf r}$, at time $t$  with a wavevector ${\bf k}$ and a growth
rate $s$:
${\bf x}={\bf x}_0+{\bf x}_1\exp[st+i{\bf kr}]$, where ${\bf x}_0$ is a steady state solution.
The equation for the perturbation ${\bf x_1}$ can formally be written in matrix
form $M {\bf x_1}= {\bf 0}$ and the possible growth rates of the perturbation
are determined from
the equation $\det(M)=0$. This is a quadratic equation in $s$ with two roots denoted
by $s^+, s^-$ with
 ${\rm Re}(s^+)\ge{\rm Re}(s^-)$. The sign of ${\rm Re}(s^+)$ gives the stability limit of the steady state homogeneous phase.


We are only able to give a complete discussion of the stability of the uniform steady states with respect to a periodic perturbation in the quasi 1-dimensional case where we do not allow for a $y$ dependence ($k_y=0$). The treatment of the
fully general 2-dimensional problem requires numerical work.
However most of the physics can be extracted from the quasi-1-dimensional
case, which we present hereafter; the numerical study
allows us to extend our conclusions to the  general 2-dimensional case.

\begin{figure}
\begin{center}\scalebox{0.3}{
\includegraphics{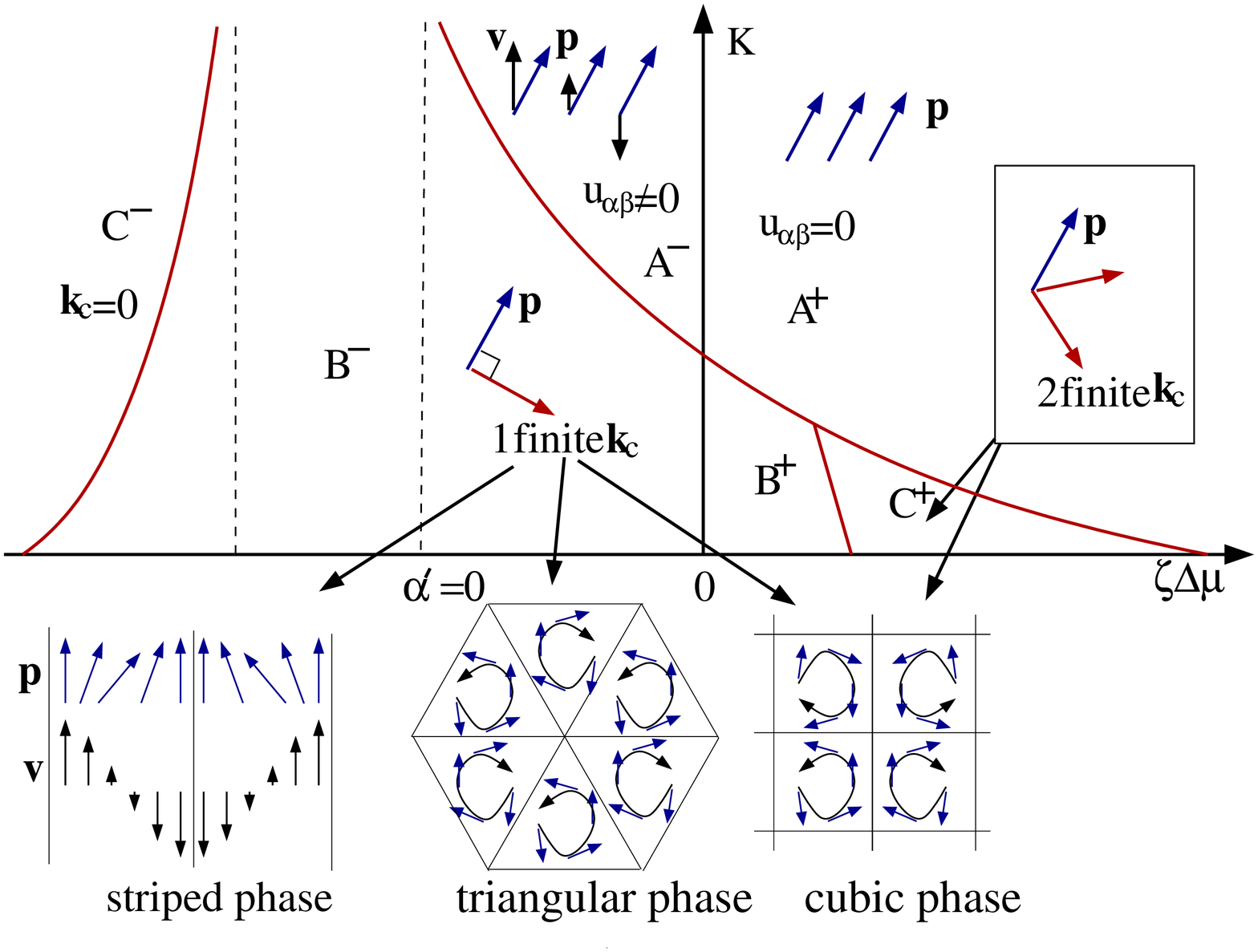}}
\caption{\label{fig1}Dynamic phase diagram of an active polar film.
The transition line between regions $B^{+}$ and $C^{+}$ is the line
$ \zeta\Delta\mu^*(K) $; the transition between regions $B^{-}$,
$B^{+}$ and $C^{+}$ and regions $A^{-}$ and $A^{+}$ is the line
$K_c(\zeta\Delta\mu)$; the transition between regions $B^{-}$ and
$C^{-}$ is the line $\Delta\mu_p(K)$}\end{center}
\end{figure}

We first analyze the stability of  the static state, fixing all the parameters of the active
gel introduced above, except for the Franck constant $K$ and the active stress
$\zeta \Delta\mu$. As there is no flow in the system,
there is one single
 direction, given by the polarization, which we choose as the $y$ axis
($\theta_0=\pi/2$). The numerical study of the 2-dimensional problem
reveals that the maximum of ${\rm
Re}[s^+(k_x,k_y)]$ lies on the axis $k_y=0$, for any value
$\zeta\Delta\mu\le\zeta\Delta\mu^*(K)$, with
$\zeta\Delta\mu^*(K)>0$. In this regime, it is therefore sufficient
to consider one-dimensional perturbations in the  $x$ direction.
The growth rates are then solutions of:
\begin{widetext}
\be\label{zeta+}
2\eta\gamma a s^2+s\left[(2\eta\beta+Kab\gamma)k^2+2\eta\alpha+\zeta\Delta\mu\gamma a(\nu+1)\right]
+bK\beta k^4+\left[b(K\alpha-w^2)+\zeta\Delta\mu \beta(\nu+1)\right]k^2+{ \zeta}\Delta\mu \alpha(\nu+1)=0
\end{equation}
\end{widetext}
where  we have redefined $k_x\equiv k$ and where $a $ and $b$ are dimensionless functions of the parameters: $a=2\frac{\eta}{\gamma}+{\bar \nu}(\nu+{\bar \nu})$ and $ b=2\frac{\eta}{\gamma}+\frac{(1+\nu)^2}{2}$.
It is useful to analyze first the passive case
where $\zeta\Delta\mu=0$. If $K\ge w^2/\alpha$, ${\rm Re}(s^+)$ is
maximum and negative for $k=0$ and the  active gel is stable. If
$K< w^2/\alpha$, one can check that ${\rm Re}(s^+)$ is maximum and
positive when  $k=k_{c}\not=0$. The uniformly ordered phase is
locally unstable with respect to a finite  wavelength
longitudinal mode transverse to the ordering direction. It follows
that in this region of the phase diagram there necessarily exists
an ordered modulated phase characterized by a periodic polarization.
The precise symmetry of this passive phase, (striped, hexagonal or
cubic phase), has been discussed in \cite{blank,pelc} for the
case of ferroelectric nematic liquid crystals. We now consider the effect of
the activity on this instability. For ${ \zeta}\Delta\mu<0$, the static state is
unstable at zero wave vector (both growth rates are real and
$s^+s^-<0$), with respect to the flowing state, as found in Ref.\cite{voit}.
For ${\zeta}\Delta\mu>0$,
the uniform state is unstable with respect to a finite  wavelength longitudinal
 mode, transverse to the polarization direction when the Franck constant
is small enough  $K<K_c({ \zeta}\Delta\mu)$. The function
$K_c({\zeta}\Delta\mu)$  can be analytically calculated from
Eq.\ref{zeta+} in this regime and is plotted in Fig.\ref{fig1}. For
$\zeta\Delta\mu>\zeta\Delta\mu^*(K)>0$, a 2-dimensional numerical
analysis is required and reveals that for $K<K_c({\zeta}\Delta\mu)$
(evaluated by numerically computing the sign of ${\rm Re}(s^+)$),
there are 2 independent most unstable wavevectors ${\bf k}_{c1}$,and
$\ {\bf k}_{c2}$.

We now discuss the flowing state (${
\zeta}\Delta\mu<0$), which we believe to be relevant for
cytoskeleton dynamics. (If ${
\zeta}\Delta\mu>0$, the flowing state is unstable at any wave vector).
The numerical study of a 2-dimensional perturbation
shows that the maximum of the growth rate ${\rm
Re}[s^+(k_x,k_y)]$ is obtained for a wavevector $\bf k$ perpendicular to the polarization,
for any value $\zeta\Delta\mu\le 0$. It is therefore sufficient
to consider an effective 1-dimensional problem where spatial variations
are allowed only along the unstable direction orthogonal to ${\bf p}$. The growth
rates are in this case solutions of :
 \begin{widetext}
\be\label{s(k)}
2\eta\gamma a s^2+s\left[(2\eta\beta+Kab\gamma)k^2+2\eta\alpha'\right]
+bK\beta k^4+b(K\alpha-w^2)k^2+ikdw\zeta\Delta\mu -{ \zeta}\Delta\mu \alpha(\nu+1)=0
\end{equation}
\end{widetext}
where $\alpha'$ is an effective   compressibility,
and $d$ a dimensionless function of the parameters:
$d=2\frac{\eta}{\gamma}+(\nu+1)/2+ {\bar \nu}$ and
$\alpha'=\alpha+d\gamma\zeta\Delta\mu(\nu+1)/2$.

 For  $K<K_c({
\zeta}\Delta\mu)$ (evaluated by numerically computing the sign of
${\rm Re}(s^+)$ from Eq.\ref{s(k)}), the uniform state is unstable
with respect to a finite  wavelength mode. The value of the critical
Franck constant $K_c$ diverges for an active stress
${\zeta}\Delta\mu^c$ such that $\alpha'({\zeta}\Delta\mu^c)=0$. At
lower values of the active stress, the flowing state is always
unstable. The most unstable wave vector $|{\bf k}_c|$ decreases with
${\zeta}\Delta\mu$ and vanishes if the active stress is smaller than
a critical value $\zeta\Delta\mu_p(K)$.  Note that in this regime of
negative active stress, ${\rm Im}(s^+)\not=0$ and the instability is
oscillatory, as opposed  to the regime ${\zeta}\Delta\mu>0$.

The results of the stability analysis are summarized in the phase diagram of Fig.\ref{fig1}.
A full analysis of the nonlinearities is necessary to predict the symmetry of the dynamic
equilibrium states. This would require an exhaustive study of all the possible quadratic
and cubic terms in the perturbations in the equations of motion, which seems out of reach.
We can nevertheless infer qualitatively the structure of the different phases.

In regions B$^-$ and B$^+$ of the phase diagram,
the instability occurs at finite wavevector. It is similar to the instability of the passive system
described in \cite{pelc,blank}: it  favors splay in the system, which can appear either
in a striped or "lattice"
(cubic, hexagonal, or triangular) phase.  Because of the outbreak of polarization gradients,
the strain rate tensor cannot vanish and the system flows. The continuity
of the  velocity field imposes extra constraints and does not allow for the formation
of domain walls. This, together with the results of \cite{krus04} suggests that
in the lattice phases, each elementary cell contains a spiral--like structure with
a rotating flow. This structure is not compatible with the hexagonal symmetry :
the only possible lattice phases have therefore cubic or
triangular symmetries sketched on Fig.\ref{fig1}.

In region C$^-$, the system is unstable for ${\bf k}=0$ and thus its
behavior depends on boundary conditions. We expect a macroscopic
phase separation since the effective compressibility $\alpha'$
becomes negative. In region C$^+$,  the existence of 2 independent
unstable modes  suggests that an oblique phase prevails, even if non
linear terms could a priori select only one of the unstable modes.

 Our analysis has been presented in terms of the actin cytoskeleton,
but it is general enough to be applied to {\it any} visco-elastic,
polar, and active material. We now briefly discuss two other
examples. One should keep in mind that our model does not include
noise, be it thermal or intrinsic (due to the stochastic activity of
motors, actin polymerization...) and that the predicted ordered
phases could be disordered in the presence of noise.

Systematic quantitative data describing the phases of in vitro
acto-myosin solutions as a function of the contractility (associated
to  changes either in myosin II and/or ATP concentrations) and
therefore of $\zeta\Delta\mu$ do not seem to be available yet.
However, both disordered vortex flowing phases \cite{menon,kaes},
which could correspond to region B$^-$,  and phase separations
(superprecipitation, \cite{japa}, corresponding to region C$^-$)
are observed in experiments or obtained by numerical simulations.
Microtubules and kinesins solutions also belong to the class of
materials described by our model. They are viscous, polar, and
active. The phase diagram obtained in this paper  qualitatively
agrees with the phases observed in vitro for microtubules and
kinesin motors in two dimensions \cite{nede97}: when the motor
concentration is increased, a flowing phase of spirals appears (B);
for higher motor concentrations, the spirals become growing asters
which progressively separate (C$^-$); for even higher concentrations
microtubules bundles are formed (C$^-$).

A similar behavior is observed in bacterial colonies : experiments
on 2-dimensional colonies of B. Subtilis\cite{gold} show a bacterial
flow composed of rotating whirls of swimming bacteria. This
bacterial turbulence could be described by a disordered version of
the cubic phase (B$^+$, C$^+$ or B$^-$) predicted by our model.
Indeed these bacteria are rod--like shaped, oriented by their
flagella, and therefore polar; in a coarse--grained picture,  a
colony can be viewed as a viscoelastic gel; last, they consume
chemical energy ($0_2$) and hence are active. The alternative model
of \cite{simh02} also predicts an intrinsic flow instability for
self propelled particle assemblies. This instability differs from
the one discussed here in that it has no threshold. The main
differences with our description are first that  our model in this
one fluid version cannot impose a spontaneous velocity of the gel
with respect to the background fluid; second, our model describes a
compressible gel, this feature being crucial to trigger the
instabilities. A thorough comparison between the two theories needs
a two fluids description and is underway.

\acknowledgments We are grateful to F. Amblard (Institut Curie), D.
R. Nelson (Harvard University),  D. Riveline (Grenoble) and S.
Ramaswamy (Bangalore) for very useful discussions.


\begin{thebibliography}{0}
\bibitem{simh02}
R.A. Simha and S. Ramaswamy, {\it Phys. Rev. Lett.} {\bf 89}, 058101 (2002);
Y. Hatwalne, S. Ramaswamy, M. Rao and R.A. Simha, {\it Phys. Rev. Lett.} {\bf 92}, 118101 (2004)

\bibitem{gold} C. Dombrowski, L. Cisneros, S. Chatkaew, R. E. Goldstein, and J. O. Kessler {\it Phys. Rev. Lett.}  {\bf 93}, 098103 (2004)

\bibitem{albe02} B. Alberts {\it et al.},
{\it Molecular Biology of the Cell}  4th ed. (Garland, New York, 2002).

\bibitem{howa01}
J. Howard,
{\it Mechanics of Motor Proteins and the Cytoskeleton} (Sinauer Associates, Inc., Sunderland, 2001).

\bibitem{nede97}
F.J. N\'ed\'elec {\it et al.},
{\it Nature}  {\bf 389}, 305 (1997).

\bibitem{krus00}
K. Kruse and F. J\"ulicher,
{\it Phys. Rev. Lett.} {\bf 85}, 1778 (2000); K. Kruse, S. Camalet, and F. J\"ulicher,
{\it Phys. Rev. Lett.} {\bf 87}, 138101 (2001).

\bibitem{lee01}
H.Y. Lee and M. Kardar,
{\it Phys. Rev. E} {\bf 64}, 056113 (2001).

\bibitem{kim03}
J. Kim {\it et al.},
{\it J. Korean Phys. Soc.} {\bf 42}, 162 (2003).

\bibitem{live03}
T.B. Liverpool and M.C. Marchetti,
{\it Phys. Rev. Lett.} {\bf 90}, 138102 (2003).

\bibitem{japa} T. Sekine and M. Yamaguchi, {\it J. Biochem} {\bf 59(2)}, 195 (1966); Y. Nonomura and S.J. Ebashi, {\it Mechanochem. Cell Motil.} {\bf 3(1)}, 1 (1974)


\bibitem{menon}
S. Sankararaman, G.I. Menon and P.B. Sunil Kumar,
{\it Phys. Rev. E} {\bf 70}, 031905 (2004),
M.C. Aronson {\it Unpublished}





\bibitem{krus04} K. Kruse, J.F. Joanny, F. J\"ulicher, J. Prost and K. Sekimoto,
Phys. Rev. Lett {\bf 92}, 078101 (2004) and Eur. Phys. J. E {\bf
16}, 5 (2005).

\bibitem{voit} R. Voituriez, J.F. Joanny and J. Prost,
{\it Eur. Phys. Lett.} {\bf 70}, 3 (2005).

\bibitem{pelc} G.A. Hinshaw Jr., R.G. Petschek and R.A. Pelcovits, {\it Phys. Rev. Lett.} {\bf 60}, 18 (1988).

\bibitem{blank} D. Blankschtein and R.M. Hornreich {\it Phys. Rev. B} {\bf 32}, 3214 (1985).

\bibitem{dege93}
P.G. De Gennes and J. Prost,
{\it The Physics of Liquid Crystals} (Clarendon Press, Oxford 1993).

\bibitem{kaes} K. Takiguchi, J. Biochem {\bf 109}, 502 (1991)

\bibitem{sack} J. Uhde, M. Keller, E. Sackmann, A. Parmeggiani, and E. Frey
{\it Phys. Rev. Lett.} {\bf 93}, 268101 (2004)

\bibitem{lambda} We also considered the case where $\zeta$ and $\bar \zeta$ vanish and
$\lambda$ is varied.  There is no phase separation in this case.









\end{thebibliography}
\end{document}